\documentclass[PRB,twocolumn,showpacs,preprintnumbers,superscriptaddress]{revtex4}
\usepackage{amsfonts}
\usepackage{txfonts}
\usepackage{graphicx}
\usepackage{bm}
\usepackage{subfigure}

\begin{document}

\title{Spin and mass imbalance in a mixture of two species of fermionic
atoms in a 1D optical lattice}
\author{Wen-Long Lu}
\affiliation{Department of Physics and ITP, The Chinese University of Hong Kong, Hong
Kong, China}
\author{Zhi-Guo Wang}
\affiliation{Department of Physics and ITP, The Chinese University of Hong Kong, Hong
Kong, China}
\affiliation{Department of Physics, Tongji University, Shanghai 200092, China}
\author{Shi-Jian Gu}
\email{sjgu@phy.cuhk.edu.hk}
\affiliation{Department of Physics and ITP, The Chinese University of Hong Kong, Hong
Kong, China}
\author{Hai-Qing Lin}
\affiliation{Department of Physics and ITP, The Chinese University of Hong Kong, Hong
Kong, China}

\begin{abstract}
In this paper, we study the role of both "spin"(species) and mass imbalance in
a mixture of two species of fermionic atoms with attractive interaction in an
one-dimensional optical lattice. Using the bosonization approach, quantum phase
transitions between a liquid phase and phase separated states are studied under
various conditions of interaction, spin imbalance, and mass imbalance. We find
that, in the phase-separated region, there exists two kinds of phase separation
and a special quantum phase transition might exist between them in the large
$U$ limit. On the other hand, the singlet superconducting correlation dominates
in the liquid phase. The pairing behavior has been also demonstrated that there
is oscillating behavior in real space. We find both the spin and mass imbalance
are in favor of the formation of Fulde-Ferrell-Larkin-Ovchinnikov state.
\end{abstract}

\date{\today }
\maketitle

%\pacs{73.23.Ra, 73.23.-b, 05.60.Gg, 72.10.-d}

%71.10.Fd Lattice fermion models (Hubbard model, etc.)

%03.75.Mn Multicomponent condensates; spinor condensates

%05.70.Jk Critical point phenomena

\section{Introduction}

Ultracold atoms trapped in optical lattices have been increasingly
used to simulate the rich physics in strongly correlated condensed
matter systems. In particular, the Fulde-Ferrell-Larkin-Ovchinnikov
(FFLO) state \cite{PFulde, ALarkin} in a magnetized superconductor,
in which spin imbalance of electrons leads to Cooper pair formation
and condensate in states with nonzero momentum (inhomogeneous
distribution in real space), was proposed four decades ago. However,
the observation of the FFLO state in solids has been proven to be
extremely difficult due to the Meissner effect, and was only
achieved recently in a heavy fermion system \cite{HARadovan}.

Experimentally, two hyperfine states of ultracold fermionic atoms
play the roles of up and down spins. Their population could be
controlled by using radio-frequency field, such that the hope of
observing the FFLO state in cold atomic systems has been renewed
recently \cite{MWZwierlein, YShin, GBPartridge}. Since the
dimensionality of the cold atomic systems can be easily tuned, and
indeed cold atoms have already been successfully trapped in
one-dimensional (1D) waveguide, it seems natural to consider these
nonhomogeneous pairing behaviors in this low dimensional systems.

So far, much attention has been paid to 1D spin-polarized fermionic
systems with attractive interaction by using different methods and
techniques, such as Bethe-ansatz \cite{GOrso}, density-matrix
renormalization-group \cite{AEFeiguin}, Quantum Monte Carlo
\cite{GGBatrouni}, and bosonization \cite{KYang}. All of these
studies are based on approximating the system by the 1D Hubbard
model in either uniformly distributed or harmonic trap cases
respectively. In uniformly distributed case, the
dominant order in the ground state is singlet superconducting (SS) \cite%
{GGBatrouni}, and existence of FFLO state \cite{GOrso, AEFeiguin,
GGBatrouni, KYang}, which is oscillating in real space and peak in momentum
space (Non-zero momentum of cooper pairs). As polarization increasing
gradually, phase separation will happen gradually in the Harmonic trap, like
FFLO \& fully paired wings, FFLO \& polarized wings and
Chandrasekhar-Clogston limit \cite{GOrso} .

Recently, Taglieber \emph{et al} \cite{MTaglieber} have successfully trapped
a quantum degenerate Fermi-Fermi mixture, i.e. $^6$Li and $^{40}\text{%
K}$, by evaporating cooled bosonic $^{87}\text{Rb}$ gas. The advance raises a
new and interesting question. What is the influence of mass difference to the
spin polarized fermionic system or does it become easier to observe the FFLO
state in such a system. In Ref. \cite{MACazalilla}, the influence of mass
difference has been discussed with $N_\uparrow = N_\downarrow$. They found that
there is a phase transition between SS and charge density wave
(CDW) in the negative $U$ case, and phase separation exists in the positive $%
U$ case. The bosonization study \cite{ZGWang} showed that the phase boundary
between the density-wave phase and phase separation scales like $U^2$ in the
weak coupling region. The results are consistent with previous numerical
studies \cite{SJGuRFan} by the exact diagonalization and density-matrix
renormalization group technique.

The main purpose of this paper is to study the role of both
``spin"(species) and mass imbalance in a mixture of two species of
fermionic atoms with attractive interaction in optical lattices. We
use the 1D asymmetric Hubbard model (AHM) as an effective model to
describe such a mixture in the optical lattice, and then study the
phase separation, dominant order and pairing behavior in the context
of different spin populations and negative $U$ region using the
bosonization approach. We find that if the system is partially
polarized, there can exist two different phase-separated states, and
the FFLO state might be more stable in the presence of mass
imbalance.

The paper is organized as follows. In section \ref{sec:res}, we obtain the
bosonized form of the 1D asymmetric Hubbard model, then simplify it under
conditions of away from half filling $n < 1$ and $U<0$. In section
\ref{sec:ps}, we study the ground-state phase diagram, and the effects of spin
imbalance, and mass imbalance. In section \ref{sec:do}, we discuss dominant
order and pairing behavior in our interested system. Finally, we summarize our
results in section \ref{sec:sum}.

\section{THE ASYMMETRIC HUBBARD MODEL AND ITS BOSONIZED FORM}

\label{sec:res}

Here we consider a mixture of two species of fermionic atoms loaded in a 1D
optical lattice. In experiments, Such an optical lattice potential can be
written as
\begin{equation}
V(x,y,z)=V_{0}\sin ^{2}(kx)+V_{\perp }[\sin ^{2}(ky)+\sin ^{2}(kz)]
\end{equation}%
where $V_{0}(V_{\perp })=v_{0}(v_{\perp })E_{R}$ in unit of the recoil energy
$E_{R}=\hbar ^{2}k^{2}/2m$. If $v_{\perp }\gg v_{0}$, the hoping process in the
$yz$ plane is frozen; while along $x$ direction, the hopping integrals depend
on the mass of atoms. So the system is quasi
one-dimensional. Without loss of generality, we use \textquotedblleft spin" $%
\sigma =\uparrow,\downarrow $ to denote the type of atoms. For sufficiently
low temperatures, the atoms will be confined to the lowest Bloch band, then
the system can be described by the 1D AHM, whose Hamiltonian reads \cite%
{MACazalilla,ZGWang,SJGuRFan}
\begin{equation}
\mathcal{H}=-\sum_{\sigma, j}t_{\sigma }\left( c_{j\sigma }^{\dag }c_{j+1\sigma
}+h.c.\right) +U\sum_{j}n_{j\uparrow }n_{j\downarrow }.
\end{equation}%
where $c_{j\sigma }^{\dag }$ ($c_{j\sigma }$) are fermion creation
(annihilation) operators at site $j(j=1,\dots,L)$, $n_{j\sigma
}=c_{j\sigma }^{\dag }c_{j\sigma }$ , and $U$ the on-site
interaction between two species of atoms. In the following we set
$t_\uparrow$ as unit, $N_\sigma=\sum_j n_{j\sigma}$, band filling
$n=(N_\uparrow+N_\downarrow)/L$, mass ratio
$t=t_\downarrow/t_\uparrow=m_\uparrow/m_\downarrow$, mass imbalance
$z=(t_{\uparrow }-t_{\downarrow })/(t_{\uparrow }+t_{\downarrow })$,
and polarization
$P={|N_\uparrow-N_\downarrow|}/(N_\uparrow+N_\downarrow)$.

In the standard bosonization method \cite{VJEmery, DGShelton}, the AHM can be
expressed in terms of canonical Bose fields and their dual counterparts as
\begin{eqnarray}
\mathcal{H_{B}} &=&\frac{v_{c}}{2}\int dx\left[ \frac{1}{K_{c}}(\partial _{x}%
{\phi _{c}})^{2}+K_{c}\pi _{c}^{2}\right]  \nonumber \\
&+&\frac{v_{s}}{2}\int dx\left[ \frac{1}{K_{s}}(\partial _{x}{\phi _{s}}%
)^{2}+K_{s}\pi _{s}^{2}\right]  \nonumber \\
&+&\frac{U}{2\pi ^{2}a}\int dx\cos \left[ \sqrt{8\pi }\phi
_{c}+2(k_{F\uparrow }+k_{F\downarrow })x\right]  \nonumber \\
&+&\frac{U}{2\pi ^{2}a}\int dx\cos \left[ \sqrt{8\pi }\phi
_{s}+2(k_{F\uparrow }-k_{F\downarrow })x\right]  \nonumber \\
&+&\Delta {v}\int dx\left[ \pi _{c}\pi _{s}+\partial _{x}{\phi
_{c}}\partial _{x}{\phi _{s}}\right],
\end{eqnarray}%
where all parameters take the following forms
\begin{widetext}
\begin{equation}
v_c = a\sqrt{t_{\uparrow}\sin(k_{F\uparrow}a)+t_{\downarrow}
\sin(k_{F\downarrow}a)\left[t_{\uparrow}\sin(k_{F
\uparrow}a)+t_{\downarrow}\sin(k_{F\downarrow}a)+\frac{U}{2\pi}\right]},\\
\end{equation}
\begin{equation}
v_s = a\sqrt{t_{\uparrow}\sin(k_{F\uparrow}a)+t_{\downarrow}
\sin(k_{F\downarrow}a) \left[t_{\uparrow}\sin(k_{F
\uparrow}a)+t_{\downarrow}\sin(k_{F\downarrow}a)-\frac{U}{2\pi}\right]},\\
\end{equation}
\end{widetext}%
\begin{equation}
\frac{1}{K_{c}}=\sqrt{1+\frac{U}{2\pi \lbrack t_{\uparrow }\sin
(k_{F\uparrow }a)+t_{\downarrow }\sin (k_{F\downarrow }a)]}}, \\
\end{equation}%
\begin{equation}
\frac{1}{K_{s}}=\sqrt{1-\frac{U}{2\pi \lbrack t_{\uparrow }\sin
(k_{F\uparrow }a)+t_{\downarrow }\sin (k_{F\downarrow }a)]}}, \\
\end{equation}%
\begin{equation}
\Delta {v}=a[t_{\uparrow }\sin (k_{F\uparrow }a)-t_{\downarrow }\sin
(k_{F\downarrow }a)].
\end{equation}%
Here the bosonic fields $\phi _{c}$ and $\phi _{s}$ characterize the
charge and spin degree of freedom,respectively. $k_{F\uparrow }$ and
$k_{F\downarrow }$ are the Fermi wavevectors for up- and down-spin
atoms, which are determined by number density of each component, and
$a$ is the lattice constant. $v_{c,s}$ are the propagation
velocities of the charge
and spin collective modes of the decoupled systems ($\Delta {v}=0$), and $%
K_{c,s}$ are the stiffness constants.

There are two oscillating terms, Umklapp and backward terms, in our low
energy effective Hamiltonian. If $k_{F\uparrow }+k_{F\downarrow }\neq \pi /a$
or $k_{F\uparrow }\neq k_{F\downarrow }$, they will vanish after performing
such integrals. Physically, quasi-momentum conservation laws do not hold in
the low energy region for these two processes. Even if both of them survive,
it does not mean that they will contribute significantly in the long
wavelength scale. According to renormalization-group analysis \cite{VJEmery}%
, Umklapp term $\cos (\sqrt{8\pi }\phi _{c})$ contributes effectively only
for $U>0$ case, there will be a gap in the charge excitation spectrum. On
the other hand, backward term $\cos (\sqrt{8\pi }\phi _{s})$ contributes
effectively only for $U<0$ case there will be a gap in spin excitation
spectrum.

Therefore, both of them will disappear in our systems, i.e., spin imbalance
and attractive on-site interaction $U<0$. After do one loop approximation,
the effective Hamiltonian can be written as
\begin{eqnarray}
\mathcal{H}_{eff} &=&\frac{v_c}{2}\int dx \left[\frac{1}{K_c}
\left(\partial_x{\phi_c}\right)^2+K_c\pi_c^2\right]  \nonumber \\
&+&\frac{v_s}{2}\int dx \left[\frac{1}{K_s} \left(\partial_x{\phi_s}%
\right)^2+K_s\pi_s^2\right]  \nonumber \\
&+&\Delta{v}\int dx \left[\pi_c \pi_s+\partial_x{\phi_c}\partial_x{\phi_s}%
\right] .  \label{eq:H_eff}
\end{eqnarray}

Here we want to emphasize that the coupling constant $\Delta{v}$
depends on both spin polarization and mass difference. Unlike the
Hubbard model, here $N_\uparrow > N_\downarrow$ and $N_\uparrow <
N_\downarrow$ will be related to different cases because up- and
down-spin atoms can be distinguished from each other due to mass
difference. In the following, we try to study the consequences of
mass imbalance and spin imbalance for phase separation and dominant
orders.

\section{Phase separation}

\label{sec:ps}

The quadratic effective Hamiltonian [Eq. (\ref{eq:H_eff})] can be diagonalized
in terms of two new fields $\phi_\pm$ which are combinations of spin and charge
degrees of freedom. The corresponding velocities have been obtained
\begin{widetext}
\begin{equation}
v^2_{+,-}=\frac{v^2_c+v^2_s}{2}+\Delta{v}^2\pm\sqrt{\left(\frac{v^2_c-v^2_s}{2}\right)^2+\Delta{v}^2
\left[v^2_c+v^2_s+v_c v_s\left(K_c K_s +\frac{1}{K_c
K_s}\right)\right]}
\end{equation}
\end{widetext}
As $\Delta{v}\rightarrow{0}$, $v_{+}\rightarrow \max (v_c, v_s)$, $%
v_{-}\rightarrow \min (v_c, v_s)$, and here $v_c<v_s$. As $\Delta{v}$
increases, $v_{-}$ decreases until it vanishes at the points:
\begin{eqnarray}
\Delta{v}_1^2 &=& v_c v_s K_c K_s \\
\Delta{v}_2^2 &=& v_c v_s \frac{1}{K_c K_s}.
\end{eqnarray}
At these points, the freezing of the lower bosonic (mixture of real spin and
charge) mode is accompanied by a divergence in the charge and spin response
functions. The static charge compressibility $\kappa$ diverges at $\Delta{v}%
=\Delta{v}_1$ or $\Delta{v}=\Delta{v}_2$. It behaves as
\begin{equation}
\kappa = \kappa_0 \left [1-\frac{\Delta{v}}{\Delta{v}_{1(2)}}\right ]^{-1},
\kappa_0=\frac{2K_c}{\pi v_c}.
\end{equation}

Beyond these points, the susceptibilities become negative. This behavior of the
static response functions together with the vanishing of the collective mode
velocity indicates that the ground state becomes unstable \cite{JVoit_JDrut}
and undergoes a first-order phase transition \cite{JVoit}. The instability is
known as phase separation and has been shown to occur in the extended Hubbard
Model and in the $t-J$ model\cite{KPenc, VJEmery_Lin}. In our case, we obtain
\begin{eqnarray}
\Delta {v}_{1} &=&a(t_{\uparrow }\sin (k_{F\uparrow }a)+t_{\downarrow }\sin
(k_{F\downarrow }a)), \\
\Delta {v}_{2} &=&\pm \sqrt{\left[ a(t_{\uparrow }\sin (k_{F\uparrow
}a)+t_{\downarrow }\sin (k_{F\downarrow }a))\right] ^{2}-\frac{U^{2}}{4\pi
^{2}a^{2}}.}
\end{eqnarray}%
It is obvious that if $\Delta {v}^{2}\geq \Delta {v}_{2}^{2}$, the system is
in PS region. After doing some calculations \cite{ZGWang}, we arrive at the
condition of phase separation:
\begin{equation}
\cos \left[ \frac{(N_{\uparrow }-N_{\downarrow })\pi }{N}\right] -\cos
\left( \frac{N_{e}\pi }{N}\right) \leq \frac{U^{2}}{8\pi t_{\uparrow
}t_{\downarrow }}  \label{eq:PS}
\end{equation}%
It seems that up- and down-spin are symmetric in above expression (Bosonization
method only work in the weak coupling region). However, it is not the case in
the large $U$ limit. In the following, we will discuss these two cases in more
details.

\begin{figure}
\includegraphics[width=8.5cm]{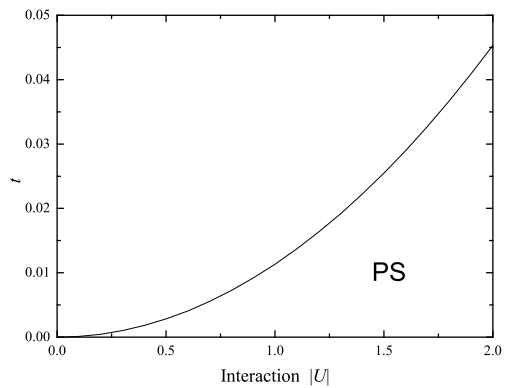}
\caption{The phase boundary between the phase separation (below the line) and
liquid phase (above the line) in the $t-|U|$ plane. Here the polarization
$P=0.5$, band filling $n=0.8$, and $U<0$.} \label{fig:PS_I}
\end{figure}

In negative $U$ case, atoms with opposite spin try to form singlet
pairs and lower the ground-state energy further. Singlet pairs can
be regarded as a kind of quasi-particle of mass $m_{\uparrow
}+m_{\downarrow }$. Generally speaking, there are three kinds of
particles, up- and down-spin atoms, and bound pairs. The reason for
phase separation is the large difference of their mass, i.e., heavy
atoms will stay together and give more space for other atoms to hop
more freely. In Fig. \ref{fig:PS_I}, we show the boundary of phase
separation in the weak coupling limit based on the Bosonization
approach for fixed polarization $P=0.5$, the more large interaction
$U$ is, the more pairs are (more unoccupied sites); and the heavier
down-atoms are, the difference of mass of these three kinds of
particles will large. Both of these two situations will lead to
phase separation easily. Here the phenomena we obtained is very
similar to observations in experiments \cite{MWZwierlein, YShin,
GBPartridge}.

\begin{figure}[h]
\includegraphics[width=8.5cm]{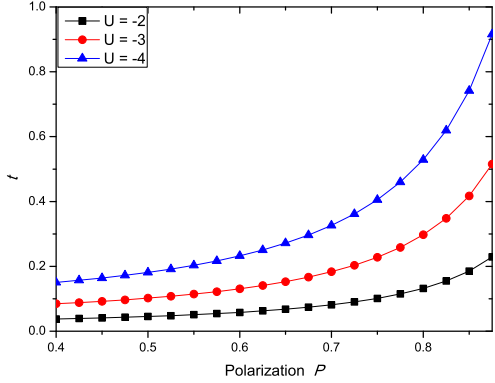}
\caption{The phase boundary between the phase separation (below the line) and
liquid phase (above the line) in the $t-P$ plane. Here band filling $ n=0.8$. }
\label{fig:ps_II.eps}
\end{figure}

\begin{figure}[tbp]
{\includegraphics[width=8.5cm]{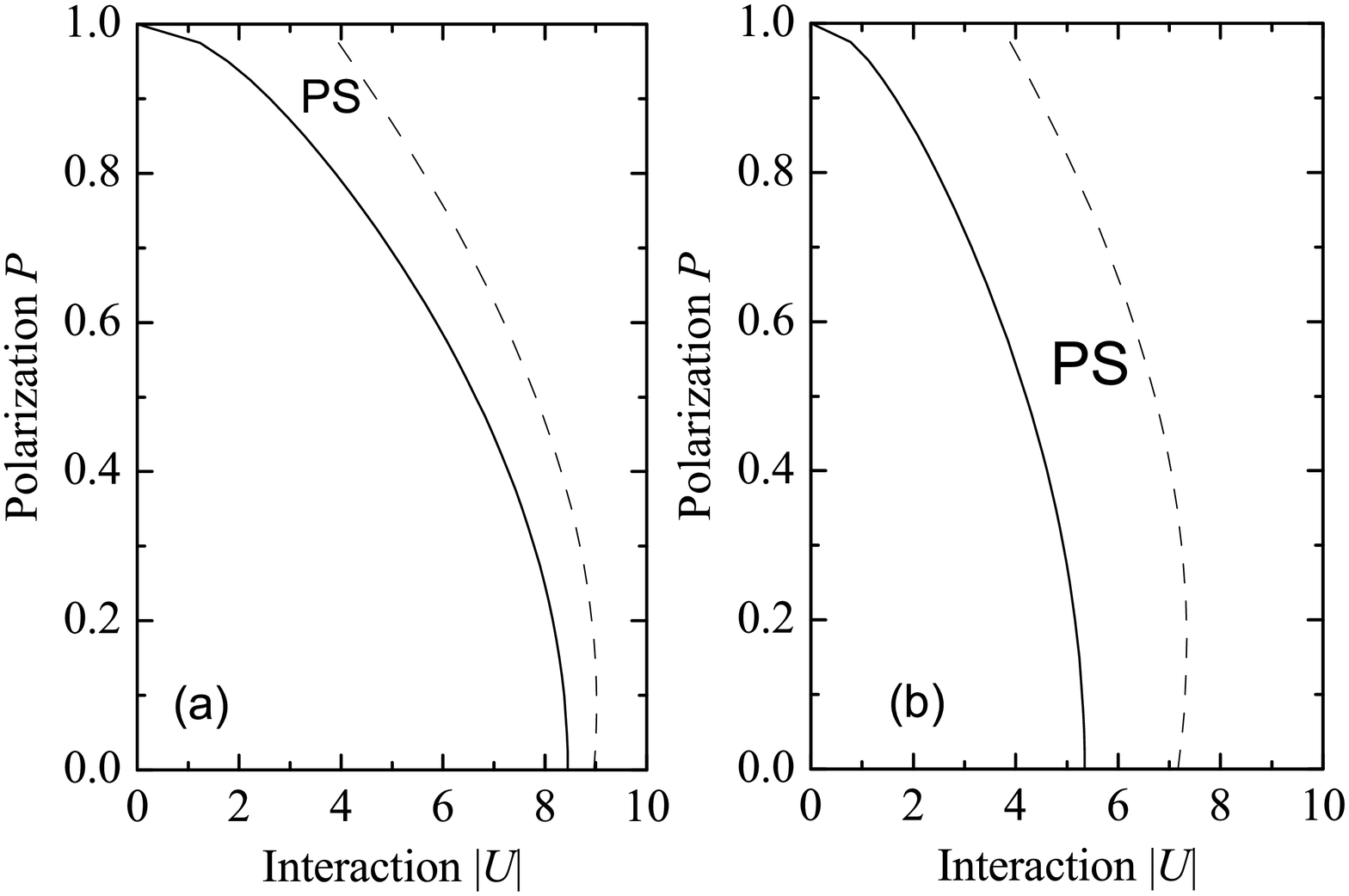}} \caption{The phase boundary (solid
line) between the phase separation (top-right region above the solid line) and
a liquid phase (bottom-left region below the solid line) in the $P-|U|$ plane
for $t=0.5$ (a) and $t=0.2$ (b). The Bosonization method is only valid on the
left side of the dashed lines. Here the band filling $n=0.8$.}
\label{fig:PS_III}
\end{figure}

The phase boundary in the $t-P$ plane under fixed interaction is obtained
similarly. If the polarization is larger under fixed band filling, there will
be less down-particles, and the system will be easily in phase separation
region composed of up-particles and bound pairs. The phase transition boundary
is shown in Fig. \ref{fig:ps_II.eps}. Two phase diagrams in the $U-P$ plane for
both cases of $t=0.5$ and $t=0.2$ are shown in Fig. \ref{fig:PS_III}. In both
cases, if the polarization $P$ is large, the phase separation state becomes
more stable. If we compare two cases, the mass imbalance will also strongly
affect the phase boundary, which will dragged to the left as the mass imbalance
becomes larger.

\section{Phase separation in strong coupling limit}
\begin{figure}[h]
\includegraphics [width=6.5cm]  {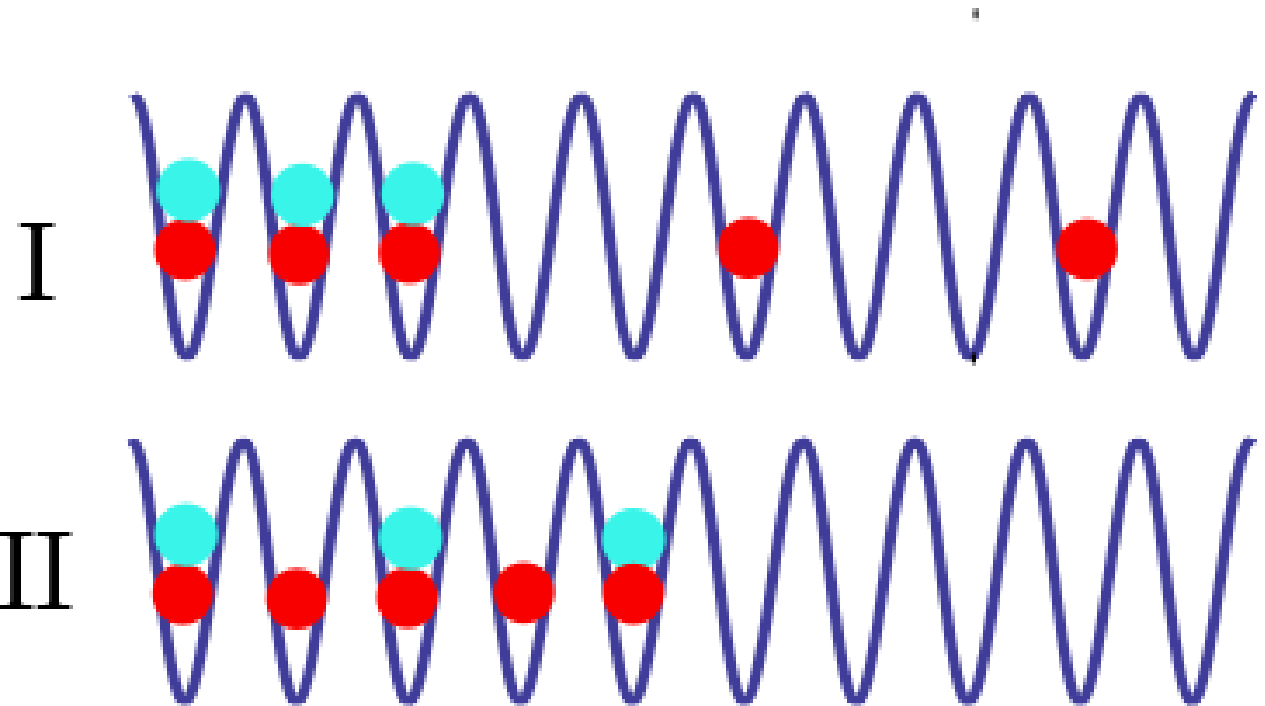}
\caption{Two possible configurations in strong coupling limit, one is pairs
staying together and unpaired particles moving; another one is dilute particles
moving in the background of dense particles that are congregated together.}
\label{fig:bestnew.eps}
\end{figure}

Bosonization approach can not provide explicit information in the
phase separation region except for indicating the phase boundary.
Therefore, we try to obtain some insights about configurations in
phase separation region by analyzing the strong coupling limit, i.e.
$U\rightarrow -\infty$. In this case, all of dilute particles are
paired with their partner. It is easy to realized that there are two
possible configurations (See Fig. \ref{fig:bestnew.eps}).

\begin{figure}[h]
\includegraphics [width=8cm] {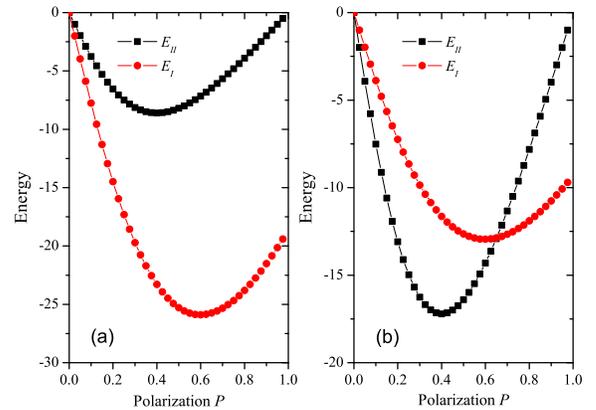}
\caption{The ground-state energy as a function of the polarization for
different dominant configurations: (a) more light atoms and (b) more heavy
atoms. Here band filling $n =0.8$, $t_{\uparrow }=1$, $t_{\downarrow }=0.5$. }
\label{fig:psconfiguration}
\end{figure}

It is obvious that there are two cases, more light particles and
more heavy particles. Now we consider the former firstly. In this
case, there are two dominant configurations for the phase
separation, i.e. the configuration I of all atom pairs staying
together and unpaired light atoms moving in free space, and the
configuration II of heavy atoms staying together and light atoms
moving in the background of heavy atoms (See Fig.
\ref{fig:bestnew.eps}). Since the interaction is infinite (only
kinetic energy--just tight bonding model), the ground-state energy of the two configurations
can be calculated by
\begin{eqnarray*}
E_{I} &=&\frac{t_{\uparrow }}{2}\left[ 1-\frac{\cos \frac{(N_{\uparrow
}-N_{\downarrow })\pi }{N-N_{\downarrow }+1}-\cos \frac{(N_{\uparrow
}-N_{\downarrow }+1)\pi }{N-N_{\downarrow }+1}}{1-\cos \frac{\pi }{%
N-N_{\downarrow }+1}}\right] , \\
E_{II} &=&\frac{t_{\downarrow }}{2}\left[ 1-\frac{\cos \frac{N_{\downarrow
}\pi }{N_{\uparrow }+1}-\cos \frac{(N_{\downarrow }+1)\pi }{N_{\uparrow }+1}%
}{1-\cos \frac{\pi }{N_{\uparrow }+1}}\right],
\end{eqnarray*}
respectively. We show the ground-state energies for both configuration, i.e.
$E(I)$ and $E(II)$, as a function of the polarization in Fig.
\ref{fig:psconfiguration}(a). From the figure, we can see that $E_{I}$ is
always smaller than $E_{II}$. Therefore, the configuration I is the ground
state. It can be understood in the following way. For light atoms, the hoping
integral is large, so they can effectively lower the ground-state energy.
While, in the configuration II, motion of heavy particles contribute less to
the ground-state energy. So the configuration I is favorable.

\begin{figure}[h]
\includegraphics [width=8.5cm] {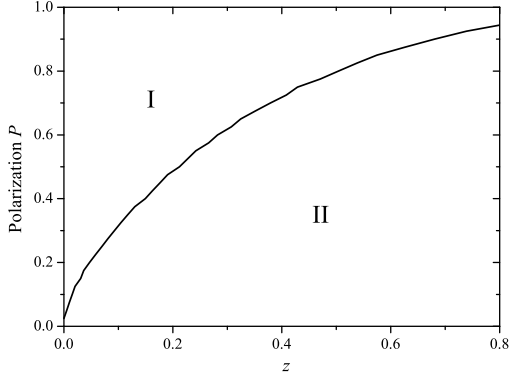}
\caption{Phase diagram for band filling $n = 0.8$ . It is configuration II
below the curve and configuration I above the curve.} \label{fig:zpolarization}
\end{figure}

On the other hand, if the heavy particles is dense, then
\[
E_{I}=\frac{t_{\downarrow }}{2}\left[ 1-\frac{\cos \frac{(N_{\downarrow
}-N_{\uparrow })\pi }{N-N_{\uparrow }+1}-\cos \frac{(N_{\downarrow
}-N_{\uparrow }+1)\pi }{N-N_{\uparrow }+1}}{1-\cos \frac{\pi }{N-N_{\uparrow
}+1}}\right] ,
\]
\[
E_{II}=\frac{t_{\uparrow }}{2}\left[ 1-\frac{\cos \frac{N_{\uparrow }\pi }{%
N_{\downarrow }+1}-\cos \frac{(N_{\uparrow }+1)\pi }{N_{\downarrow }+1}}{%
1-\cos \frac{\pi }{N_{\downarrow }+1}}\right] .
\]%
The two energies are shown in Fig. \ref{fig:psconfiguration}(b). In this case,
if $P<0.65$, the ground state is dominated by configuration II, while if
$P>0.65 $, it is by configuration I. In configuration I, more heavy particles
will contribute to the total energy, even if the value of contribution of each
heavy particle is smaller. In the configuration II, value of contribution to
total energy from each light particle is larger contribute to the total energy,
but the number of light particles is smaller. Therefore, there is a transition
point between two phase separated states. We show the phase diagram in Fig.
\ref{fig:zpolarization}. Here we would like to point out that the two phase
separated states, i.e. I and II, are qualitatively different. For the
configuration I, the pair-pair correlation function $\langle
n_{i\uparrow}n_{i\downarrow} n_{i+r\uparrow}n_{i+r\downarrow}\rangle$ has a
non-vanishing long-range behavior. So it is a true long-range order. While for
the configuration II, the correlation function decays algebraically.

\begin{figure*}[tbp]
\includegraphics[width=5.5cm]{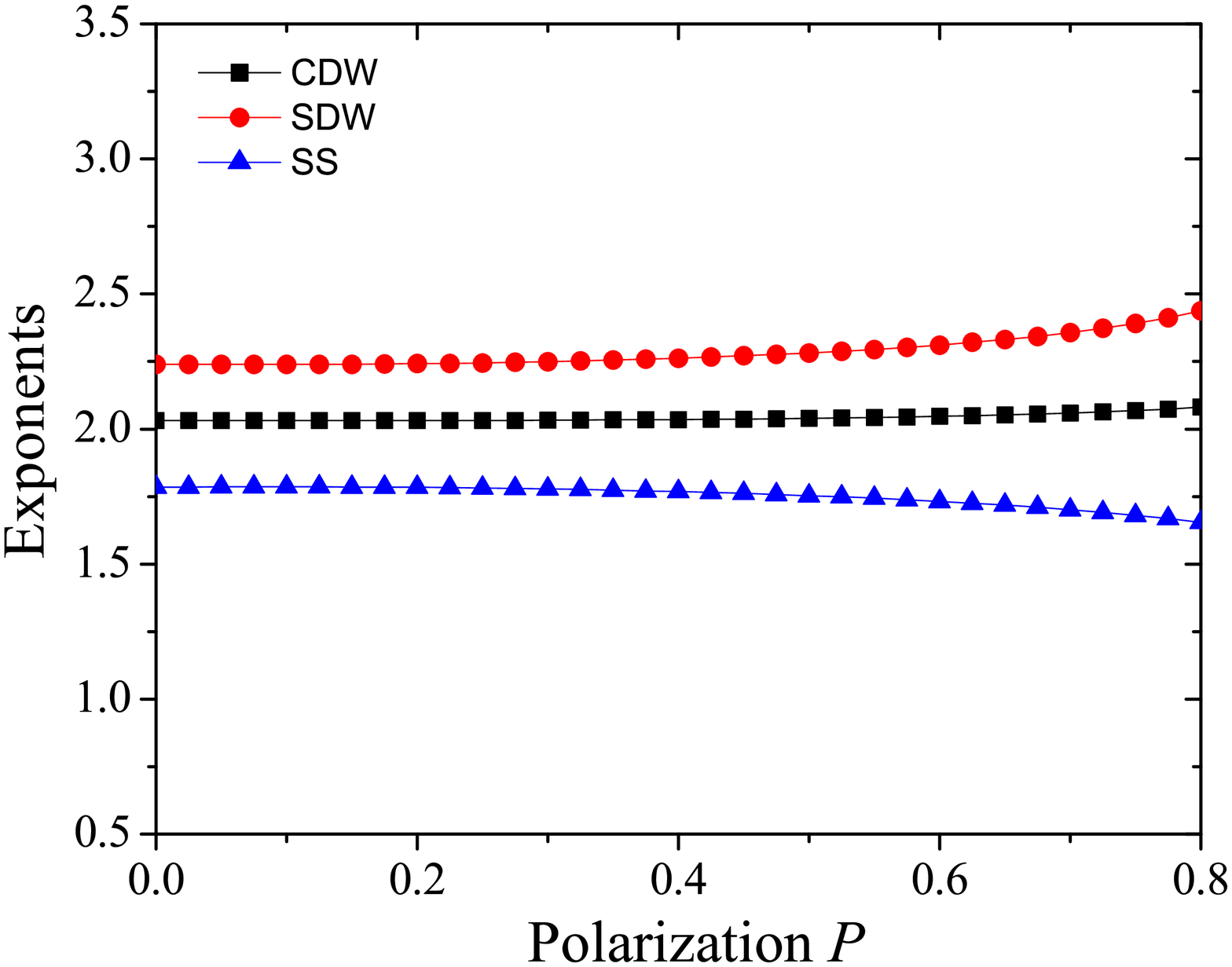}
\includegraphics[width=5.5cm]{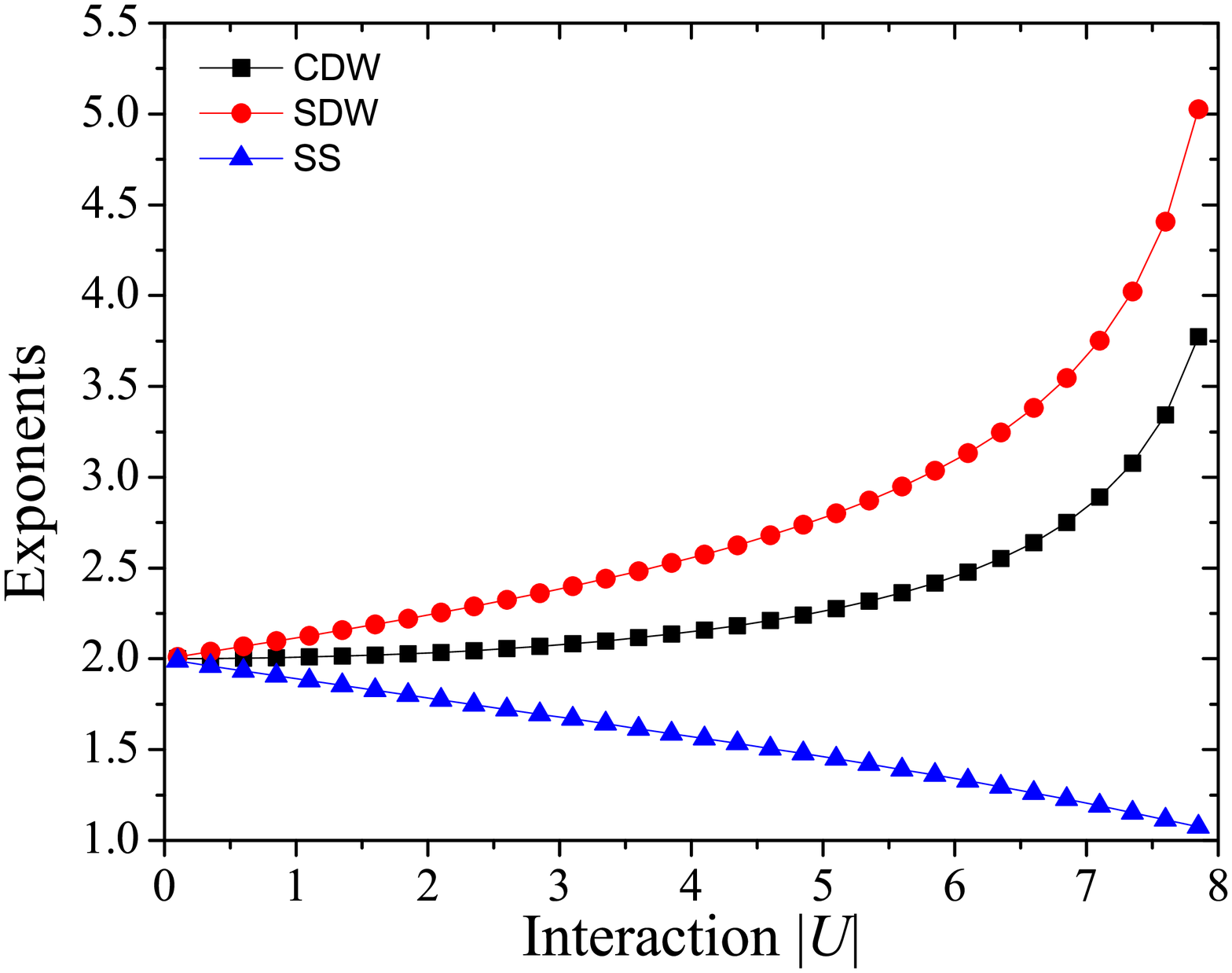}
\includegraphics[width=5.5cm]{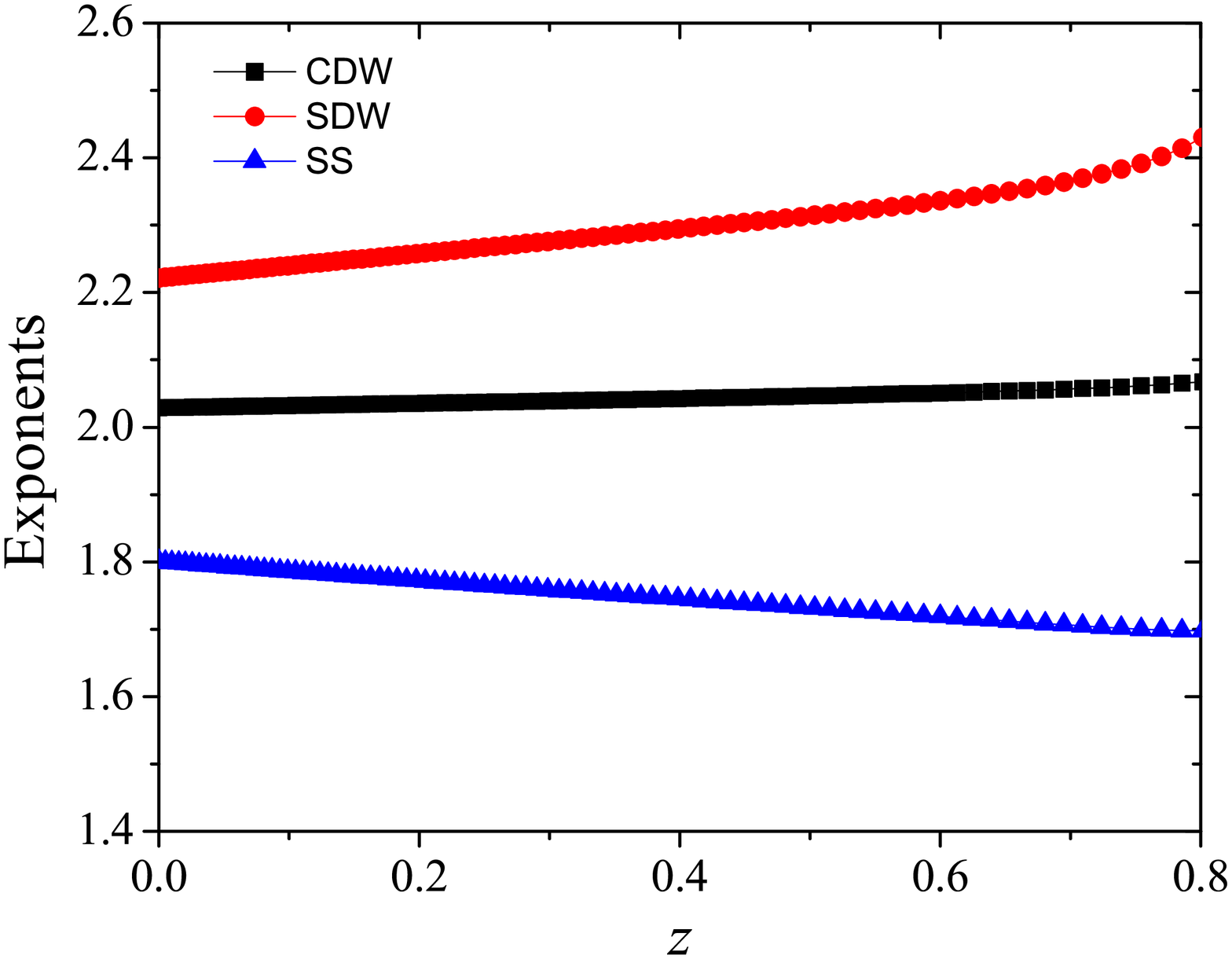}
\caption{Three exponents as a function of $P$ (left), $|U|$ (middle), and $z$
(right). Here band filling $n = 0.8$, $t_\downarrow=0.5$, $U = -2$ (for the
left one and right one). The SS order is always dominating in the liquid phase.
} \label{fig:exponents}
\end{figure*}

\section{Dominant order and pairing behavior in the liquid phase}

\label{sec:do}

As we known, the ground state is SS in standard Hubbard model with negative $%
U$. In Ref. \cite{MACazalilla}, there is phase transition from SS to CDW driven
by different mass in negative $U$ case but with $N_{\uparrow }=N_{\downarrow
}$. And now we want to find the modification of spin and mass imbalance to the
pattern in ground state. For this purpose, we need to calculate the correlation
functions related such kinds of order, like CDW, SDW, and SS. They are defined
as \cite{VJEmery}
\begin{equation}
R_{\nu }(x,x^{\prime })=\langle O_{\nu }(x)O_{\nu }^{\dag }(x^{\prime
})+h.c.\rangle
\end{equation}%
where
\begin{eqnarray}
O_{CDW}(x) &=&\sum_{\sigma }e^{-2\imath k_{F\sigma }x}\psi _{R\sigma }^{\dag
}(x)\psi _{L\sigma }(x), \\
O_{SDW}^{x,y}(x) &=&\sum_{\sigma }e^{-2\imath k_{F\sigma }x}\psi _{R\sigma
}^{\dag }(x)\psi _{L-\sigma }(x), \\
O_{SS}(x) &=&\sum_{\sigma }e^{\imath (k_{F\sigma }-k_{F-\sigma })x}\psi
_{L\sigma }^{\dag }(x)\psi _{R-\sigma }(x),
\end{eqnarray}%
for CDW, SDW, and SS, respectively. In 1D Fermi liquid, it is well known that
there is no true long-range order, and the correlation function usually has the
behavior of a power-law decay.
\begin{equation}
R_{\nu }(x,x^{\prime })\propto \frac{1}{|x-x^{\prime }|^{\alpha _{\nu }}}
\end{equation}

In Fig. \ref{fig:exponents}, we show the dependence of three correlation
exponents on $P$, $|U|$, and $z$ respectively. From the figure, we can see that
the SS correlation dominates in all three cases. Unlike the case in Ref. \cite{MACazalilla},
there is no phase transition from SS to CDW now. We believe that the system will come into phase
separation region before this transition happens.

At the same time, we can obtain oscillating behavior of pairs very easily. In our system, pair
correlation mainly comes from singlet pair correlation because of on-site
attractive interaction. Therefore, SS correlation function is just expected
now. It is obvious that pair correlation function will take the following form:
\begin{equation}
R_{SS}(x)=2\cos{[(k_{F\uparrow}-k_{F \downarrow})x]} \frac{1}{%
|x|^{\alpha_{SS}}}
\end{equation}
If and only if $N_\uparrow \neq N_\downarrow$, the pair correlation will
take the oscillation form, and different mass will modify the correlation
exponent $\alpha_{SS}$ together with spin imbalance. This state has
non-homogeneous distribution in real space, and it is called FFLO state.
From Fig. \ref{fig:exponents}, we can see that the exponent $\alpha_{SS}$ is
suppressed, hence the FFLO state becomes favorable, as the mass difference
increases. Therefore, we believe that the mass imbalance is good for the
FFLO state.

\section{Summary}

\label{sec:sum}

In this paper, from the bosonized form of the 1D asymmetric Hubbard model, we
have studied the role of spin imbalance and mass imbalance at zero temperature
under various conditions. The conditions of phase separation have been
presented, and the effects of spin imbalance, different mass and interaction
are also discussed in detail. we find that more-light-particle case is same
as the more-heavy-particle one at the weak coupling limit. And they are different
in the strong coupling limit. In the liquid phase, SS is always dominant and there is no phase transition
from SS to CDW. Finally, our results show that the mass imbalance might be
in favor of the FFLO state.

\begin{acknowledgements}
We thank helpful discussion with Shu Chen. This work is supported by the
Earmarked Grant for Research from the Research Grants Council of HKSAR, China
(Project No. CUHK 402107 and 401108).
\end{acknowledgements}

\end{document}